\documentclass[12pt]{article} 
\usepackage{amsfonts,amsmath,latexsym,amssymb,txfonts,bm} 
\usepackage{amsfonts} 
\usepackage{latexsym} 
\usepackage{amssymb} 
\usepackage[american]{babel}

\def\II{{\mathbb I}} 
\def\RR{{\mathbb R}}

\def\ZZ{{\mathbb Z}}

\def\tr{\mathrm{ tr\,}} 
\def\Tr{\mathrm{ Tr\,}}

\def\Det{\mathrm{ Det\,}}

\def\be{\begin{equation}} 
\def\ee{\end{equation}} 
\def\bea{\begin{eqnarray}} 
\def\eea{\end{eqnarray}} 
\def\bed{\begin{definition}{\ }}
\def\eed{\end{definition}}
\def\bd{\begin{description}}
\def\ed{\end{description}}
\def\bc{\begin{center}}
\def\ec{\end{center}}

\newtheorem{lemma}{Lemma}
\newtheorem{corollary}{Corollary}

\newtheorem{definition}{Definition}

\begin{document}

\begin{titlepage}
\thispagestyle{empty}
\null

\hspace*{50truemm}{\hrulefill}\par\vskip-4truemm\par
\hspace*{50truemm}{\hrulefill}\par\vskip5mm\par
\hspace*{50truemm}{{\large\sc 
New Mexico Tech {\rm (June 5, 2014) }}}\vskip4mm\par
\hspace*{50truemm}{\hrulefill}\par\vskip-4truemm\par
\hspace*{50truemm}{\hrulefill}
\par
\par
\par
\vspace{3cm}
\centerline{\huge\bf Heat Trace and}
\bigskip
\centerline{\huge\bf Functional Determinant}
\bigskip
\centerline{\huge\bf in One Dimension}
\bigskip
\bigskip
\centerline{\Large\bf Ivan G. Avramidi}
\bigskip
\centerline{\it Department of Mathematics}
\centerline{\it New Mexico Institute of Mining and Technology}
\centerline{\it Socorro, NM 87801, USA}
\centerline{\it E-mail: iavramid@nmt.edu}
\bigskip
\medskip
\vfill
{\narrower
\par

We study the spectral properties of the Laplace type operator 
on the circle. We discuss various approximations for the heat trace, 
the zeta function and the zeta-regularized determinant.
We obtain a differential equation for the heat kernel diagonal 
and a recursive system for the diagonal heat kernel coefficients,
which enables us to find closed approximate formulas for the heat trace and
the functional determinant which become exact in the limit
of infinite radius. The relation to the generalized
KdV hierarchy is discussed as well.

\par}

\vfill

\end{titlepage}


\section{Introduction}
\setcounter{equation}0

The heat kernel of elliptic partial differential operators 
is one of the most powerful tools in mathematical physics 
(see, for example,
\cite{gilkey95,hurt83,avramidi91,avramidi00b,kirsten01,vassilevich03}
and further references therein). Of special importance are the spectral
functions such as the heat trace, the zeta function and the functional determinant
that enable one to study the spectral properties of the corresponding operator.

The one-dimensional case is a very special one which exhibits an
underlying symmetry that has deep relations to such diverse areas as integrable
systems, infinite-dimensional Hamiltonian systems,
isospectrality etc
(see \cite{adler78,olmedilla81,schimming88,bilal94,avramidi95,dickey03,
avramidi00a,iliev05,iliev07a,iliev07b}, for example).
Moreover, it has been shown that one can obtain closed formulas which 
express the functional determinant in one dimension
in terms of a solution to a particular initial value problem; see, e.g., 
\cite{levit77,dreyfus78,kirsten04}.
Although the goals of our paper and the previous paper are similar
our approach is completely different. Our results are also formulated
in a completely different way. We try to obtain direct formulas
for spectral invariants in terms of the potential terms and some
new operators rather than solutions of some initial value problem.

We study the heat kernel of a Laplace type
partial differential operator 
on the circle $M=S^1$ of radius $a$.
Let ${\cal V}$ be a $N$-dimensional vector bundle over $S^1{}$ and $Q$
be a smooth Hermitian endomorphism of the bundle ${\cal V}$.
Let $L: C^\infty(\cal V)\to C^\infty({\cal V})$ be a second-order 
differential operator defined by
\be
L=-D^2+Q,
\label{4-57}
\ee
where $D=\partial_x$ denotes the derivative with respect to the local 
coordinate $x$ on $S^1{}$, 
with $0\le x\le 2\pi a$. 

The heat kernel $U(t;x,x')$ of the operator $L$ is the fundamental 
solution of the heat equation 
\be
(\partial_t+L)U(t;x,x')=0
\label{hkxx}
\ee
for $t\ge 0$ with the initial condition
\be
U(0;x,x')=\delta(x-x')
\label{icxx}
\ee

It is well known that the operator $L$ is
essentially self-adjoint in $L^2({\cal V})$ and has a discrete real 
spectrum bounded from below.
Moreover, each eigenvalue has a finite multiplicity and the corresponding
eigenvectors are smooth sections that can be chosen to form an orthonormal
basis in $L^2({\cal V})$. 
Let us denote the eigenvalues and the eigenfunctions of the operator $L$ by
$(\lambda_n, \varphi_n)_{n=1}^\infty$ where each eigenvalue 
is taken with multiplicity.
Then the heat kernel has the form
\be
U(t;x,x')=\sum_{n=1}^\infty \exp(-t\lambda_n)\varphi_n(x)\varphi^*_n(x')\,.
\ee
We note that the heat kernel diagonal $U(t;x,x)$ is a smooth self-adjoint endomorphism.

In this paper we report on various approximations
for the heat trace and functional determinant
and discuss its relation to the Korteweg-de Vries hierarchy.
Although it is heavily based upon our previous work there are many new original ideas
and results obtained in this paper.

This paper is organized as follows. In Sec. 2 we introduce the spectral invariants
such as the heat trace, the zeta function and a new very powerful invariant which
is defined in terms of the Mellin transform of the heat trace. In particular, it
immediately gives the functional determinant in one dimension.
In Sec. 3 we develop a perturbation theory in the
potential term $Q$ and compute the linear and quadratic terms in the heat trace.
In Sec. 4 we describe a scheme for the asymptotic expansion of the heat kernel
in powers of $t$ and in the Taylor series in space coordinates.
In Sec. 5 we compute the leading derivatives terms in 
the diagonal values of the heat kernel coefficients and use this to compute
the  terms linear and quadratic in the potential term in the heat trace 
and functional determinant.
In Sec. 6 we prove an algebraic lemma for the heat semigroup of 
the sum of two self-adjoint operators
and apply this lemma to obtain a differential equation directly for the
heat kernel diagonal.
In Sec. 7 we use that equation to obtain a new recursive system
for the diagonal heat kernel coefficients and obtain a closed formula
for the whole sequence of all diagonal heat kernel coefficients.
We then use this formula to obtain some closed formulas for the heat
kernel diagonal and the functional determinant. Even though
these formulas are not exact on the circle they become exact
in the limit of infinite radius. Of course, the heat trace and the 
functional determinant diverge on a noncompact space, such as the real line.
That is why, we write our formulas in terms of the circle.
In Sec. 8 we describe the bi-Hamiltonian systems and define an abstract
generalized KdV hierarchy. Then we apply this formalism to our
differential operator in one dimension and obtain the standard
KdV hierarchy, whose integrals of motion are exactly the global
heat kernel coefficients.


\section{Spectral Invariants}
\setcounter{equation}0

We will be interested in the spectral invariants of the operator $L$.
One of them, called the heat trace, is the trace of the heat kernel
and  reads
\cite{gilkey95}
\be
\Theta(t)=\Tr\exp(-tL)=\int\limits_{S^1{}} dx\,\tr\,U(t;x,x)
=\sum_{n=1}^\infty \exp(-t\lambda_n)\,.
\ee
Another important spectral invariant
is the zeta function defined by
\cite{kirsten01,avramidi00b}
\be
\zeta(s,\lambda)=\Tr (L-\lambda)^{-s}=\sum_{n=1}^\infty (\lambda_n-\lambda)^{-s},
\ee
where $\lambda$ is a {\it sufficiently large negative parameter} so that the operator
$L-\lambda$ is positive and $s$ is a complex parameter with sufficiently large 
positive real part. 
The zeta function can be expressed in terms of the Mellin transform of the
heat trace
\be
\zeta(s,\lambda)
=\frac{1}{\Gamma(s)}\int\limits_0^\infty dt\;t^{s-1}e^{t\lambda}
\Theta(t).
\ee
The zeta function enables one to define the functional
determinant as follows
\cite{kirsten01}
\be
\log\Det(L-\lambda)=-\zeta'(0,\lambda),
\ee
where $\zeta'(s,\lambda)=\partial_s\zeta(s,\lambda)$.

Next, 
we define a function $\Omega(t)$  by
\be
\Theta(t)=(4\pi t)^{-1/2}\Omega(t)\,,
\ee
and a new function $B_q(\lambda)$ 
of a complex variable $q$ as the modified Mellin transform of this function
\be
B_q(\lambda)=\frac{1}{\Gamma(-q)}\int\limits_0^\infty dt\,t^{-q-1}e^{t\lambda}
\Omega(t)\,.
\label{212}
\ee
As was shown in \cite{avramidi91},
the integral (\ref{212}) converges for ${\rm Re}\, q<0$,
and, therefore, by integrating by parts
it can be analytically continued to an entire function of $q$, that is,
for 
${\rm Re}\, q<N$,
\be
B_q(\lambda)=\frac{(-1)^N}{\Gamma(-q+N)}\int\limits_0^\infty dt\,t^{-q-1+N}\partial_t^N\left[e^{t\lambda}
\Omega(t)\right]\,.
\label{212zz}
\ee

It is also easy to see that the function $B_q(\lambda)$ is an analytic function of $\lambda$
for sufficiently large negative real part of $\lambda$, that is, for ${\rm Re} \lambda<<0$. 
Morevover, the values
of the function $B_q(\lambda)$ at non-negative integer values of $q$, that is,
$q=k=0,1,2,\dots$, 
are equal to the Taylor coefficients
of the function $\exp(t\lambda)\Omega(t)$ at $t=0$,  \cite{avramidi91},
\bea
B_k(\lambda) &=& (-\partial_t)^k \left[e^{t\lambda}\Omega(t)\right]\Big|_{t=0}\,,
\label{214xx}
\\
&=&\sum_{j=0}^k {k\choose j}(-\lambda)^{j}A_{k-j}\,,
\nonumber
\eea
where
\be
A_k=(-\partial_t)^k \Omega(t)\Big|_{t=0}\,.
\label{216xx}
\ee
Notice that for integer $q$ the functions $B_k(\lambda)$ are {\it polynomials in $\lambda$};
obviously, $B_k(0)=A_k$. However, {\it for non-integer $q$ the functions $B_q(\lambda)$ might
be singular at $\lambda=0$}.

This function enables one to express the zeta function in the form
\be
\zeta(s,\lambda)=(4\pi)^{-1/2}
\frac{\Gamma\left(s-\frac{1}{2}\right)}{\Gamma(s)}B_{\frac{1}{2}-s}(\lambda).
\ee
Now, by using the fact that $\Gamma(s)$ has a pole at $s=0$ with residue $1$, we obtain
\be
\zeta(0,\lambda)=0\,,
\ee
and a very simple formula for the 
determinant in one dimension
\be
\log\Det(L-\lambda)=B_{1/2}(\lambda)
=-\frac{1}{\sqrt{\pi}}\int\limits_0^\infty \frac{dt}{\sqrt{t}\ }\partial_t\left[e^{t\lambda}
\Omega(t)\right]\,.
\label{220xx}
\ee


We will expand the potential $Q$ in the Fourier series
\be
Q(x)=\sum_{n\in\ZZ} q_n e^{inx/a}\,,
\ee
where
\be
q_n=\frac{1}{2\pi a}\int\limits_{S^1{}}dx\; e^{-inx/a}Q(x)
\ee
and
\be
q_n^*=q_{-n}\,.
\ee


\section{Perturbation Theory}
\setcounter{equation}0

We introduce a formal small parameter $\varepsilon$ and consider the perturbation
theory for the heat trace of the operator $L=-D^2+\varepsilon Q$ which
can be obtained as a perturbation series
in powers of $\varepsilon$; we set $\varepsilon=1$ at the end.

By using the Duhamel series for the heat semigroup
\bea
U(t) &=&U_0(t)-\int\limits_0^t dv\, U_0(t-v)QU_0(v)
\nonumber\\
&&
+\int\limits_0^tdv_2\int\limits_0^{v_2}dv_1\,
U_0(t-v_2)QU_0(v_2-v_1)QU_0(v_1)+O(\varepsilon^3),
\eea
where
\be
U_0(t)=\exp(tD^2),
\ee
we get the trace
\bea
\Tr U(t) &=& \Tr U_0(t)-t \Tr QU_0(t)
\nonumber\\
&&
+\int\limits_0^t dv_2\int\limits_0^{v_2}dv_1\,
\Tr QU_0(v_2-v_1)QU_0(t-v_2+v_1)+O(\varepsilon^3).
\eea
Now, by using the formula
\be
\int\limits_0^tdv_2\int\limits_0^{v_2}dv_1\,f(v_2-v_1)
=\int\limits_0^tdv (t-v)f(v),
\ee
we obtain
\bea
\Tr U(t) &=& \Tr U_0(t)-t \Tr QU_0(t)
\nonumber\\
&&
+\int\limits_0^t dv (t-v) \Tr QU_0(v)QU_0(t-v))
+O(\varepsilon^3),
\eea
Finally, by changing the variable 
\be
v=\frac{t}{2}(1+\xi)
\ee
and using the symmetry of the integrand we get
\bea
\Tr U(t) &=& \Tr U_0(t)-t \Tr QU_0(t)
\nonumber\\
&&
+\frac{t^2}{2}\int\limits_0^1d\xi\; \Tr QU_0\left(t\frac{1+\xi}{2}\right)Q
U_0\left(t\frac{1-\xi}{2}\right)
+O(\varepsilon^3).
\eea

The heat kernel of the operator $L_0=-D^2$ is well known and has the form
\bea
U_0(t;x,x') &=& \frac{1}{2\pi a}
\sum_{n\in\ZZ}\exp\left(-\frac{t}{a^2}n^2+in\frac{(x-x')}{a}\right)
\nonumber\\
&=&
(4\pi t)^{-1/2}
\sum_{n\in\ZZ}\exp\left(-\frac{1}{4t}\left(x-x'+2\pi an\right)^2\right).
\eea
where the second form is obtained by the Poisson duality.
The heat trace then is
\be
\Tr U_0(t)=2\pi a N(4\pi t)^{-1/2}\theta\left(\frac{t}{a^2}\right),
\ee
where
\be
\theta(t) = 
\sum_{n\in\ZZ}\exp\left(-\frac{1}{t}\pi^2n^2\right)
=\frac{t^{1/2}}{\sqrt{\pi}}
\sum_{n\in\ZZ}\exp\left(-tn^2\right).
\ee
This function can be expressed in terms of the Jacobi $\theta$-function,
\be
\theta(t)=\theta_3(0,e^{-\pi^2/t}).
\ee

By using this equation we easily obtain
\bea
\Tr U(t) &=& 
(4\pi t)^{-1/2}\theta\left(\frac{t}{a^2}\right)\left(2\pi aN
-t\int\limits_{S^1} dx\,\tr Q \right)
\\
&&
+\frac{t^2}{2}
\int\limits_{S^1\times S^1} dx\,dx'\,\tr
Q(x)F(t;x,x')Q(x')
+O(\varepsilon^3),
\nonumber
\eea
where
\be
F(t;x,x')=\int\limits_0^1 d\xi\,U_0\left(t\frac{1-\xi}{2};x,x'\right)
U_0\left(t\frac{1+\xi}{2};x',x\right).
\ee

Now, by using the explicit form of the heat kernel and the Poisson duality formula
in one of the heat kernels we compute
\bea
F(t;x,x')&=&
(4\pi t)^{-1/2}
\sum_{n\in\ZZ}
\exp\left(-\frac{a^2}{t}\pi^2n^2
\right)F_n(t;x,x'),
\eea
where
\be
F_n(t;x,x')=\frac{1}{2\pi a}\sum_{k\in\ZZ}
\exp\left(ik\frac{x-x'}{a}\right)
\int\limits_0^1d\xi\,\exp\left(
-\frac{t}{a^2}\frac{1-\xi^2}{4}k^2
-ikn(1+\xi)\pi
\right).
\ee
Note that this function is the integral kernel of the operator
\bea
F_n(t)&=&
\int\limits_0^1d\xi\,\exp\left(
\frac{1-\xi^2}{4}tD^2
-n(1+\xi)\pi aD
\right),
\eea
and the function $F(t;x,x')$ is the kernel of the operator
\bea
F(t)&=&
(4\pi t)^{-1/2}
\sum_{n\in\ZZ}
\exp\left(-\frac{a^2}{t}\pi^2n^2
\right)F_n(t).
\eea

Further, we can rewrite this equation in the spectral form
\bea
\Omega(t) &=& 
\theta\left(\frac{t}{a^2}\right)
2\pi a\left(N-t\tr q_0\right)
+\pi a t^2
\sum_{k\in\ZZ}
|q_k|^2\beta_k\left(\frac{t}{a^2}\right)
+O(\varepsilon^3),
\label{317xx}
\eea
where $|q_k|^2=\tr q_kq_k^*$ and
\bea
\beta_k(t)&=&
\int\limits_0^1d\xi\,\sum_{n\in\ZZ}\exp\left(
-t\frac{1-\xi^2}{4}k^2
-\frac{1}{t}\pi^2n^2
-ikn(1+\xi)\pi
\right)
\nonumber
\\
&=&
\frac{t^{1/2}}{\sqrt{\pi}}\int\limits_0^1d\xi\,\sum_{n\in\ZZ}\exp\left\{
-t\left[n^2+\frac{(1+\xi)}{2}(k^2
-2nk)
\right]\right\}.
\eea

These formulas enable one to compute the zeta function 
and the determinant with the same accuracy, that is, up to cubic terms
in the potential $Q$. However,
we will not do it in general. Rather
we will be interested in the limit of large radius, $a\to \infty$.
We will do this in another section below by a completely different method.

Let us just note that asymptotically as $ a\to \infty$ 
\bea
\theta(t)&\sim& 1,\\
F(t)&\sim& (4\pi t)^{-1/2}
\alpha\left(-tD^2\right),
\\
\beta_k(t)&\sim& \alpha\left(tk^2\right),
\eea
where $\alpha(z)$ is a function defined by
\be
\alpha(z)=\int\limits_0^1d\xi 
\exp\left(-\frac{1-\xi^2}{4}z\right).
\label{458zz}
\ee
This is an entire function of $z$. 
By using the well known integral
\be
\int\limits_0^1d\xi\,\left(\frac{1-\xi^2}{ 4}\right)^{q}
=\frac{\Gamma(q+1)\Gamma(q+1)}{\Gamma(2q+2)}
\label{456zz}
\ee
one can obtain the power series representation of this function
\be
\alpha(z)=\sum_{k=0}^\infty \frac{k!}{(2k+1)!}(-z)^k\,.
\ee
By using either this series or by the integration by parts one can show that this function
satisfies the differential equation
\be
\left(4\partial_t+1+\frac{2}{t}\right)\alpha(t)=\frac{2}{t}\,.
\label{dea}
\ee


\section{Heat Kernel Asymptotic Expansion}
\setcounter{equation}0

It is useful to introduce various scales parametrized
by dimensionless parameters $\tau, \varepsilon$ and $\delta$
as follows. The parameter $\tau$ measures the relative radius of the circle,
\bea
\tau=\frac{t}{a^2}\,.
\eea
The parameter $\varepsilon$ measures the relative amplitude of the potential, that is,
\bea
tQ=O(\varepsilon)\,,
\eea
while
the parameter $\delta$ measures the derivatives of the potential, that is,
\bea
t^{k+1}\partial^{2k} Q=O(\delta^k\varepsilon) \,.
\eea
We assume now that $t$ is smaller than all other parameters of the same dimension, that is,
\bea
\tau\ll 1,\qquad
\varepsilon\ll 1,\qquad
\delta\ll 1\,.
\eea
Also, we consider the neighborhood of the diagonal $x=x'$, that is, we assume that
\be
x-x'=o(t^{1/2})\,.
\ee

It is well known that there is an asymptotic expansion 
as $\tau, \varepsilon, \delta \to +0$ near the diagonal of the form
\cite{gilkey95}
\be
U(t;x,x')\sim (4\pi t)^{-1/2}\exp\left\{-\frac{1}{4t}(x-x')^2\right\}
\sum\limits_{k=0}^\infty \frac{(-t)^k}{k!}a_k(x,x'),
\label{4-60}
\ee
where $a_k(x,x')$ are the so-called heat kernel coefficients.
We will denote by square brackets the diagonal values
of two-point functions, i.e.,
\be
[f]=\lim_{x\to x'}f(x,x')\,.
\ee
Then the 
asymptotic expansion of the heat kernel diagonal as $t\to 0$
is
\bea
[U(t)] &\sim& (4\pi t)^{-1/2}
\sum\limits_{k=0}^\infty \frac{(-t)^k}{k!}[a_k],
\label{4-60cx}
\eea
and, therefore, there is the corresponding asymptotics
of  the heat trace function $\Omega(t)$
\bea
\Omega(t) &\sim& 
\sum\limits_{k=0}^\infty \frac{(-t)^k}{ k!}A_k,
\label{4-60c}
\eea
where
\be
A_k=\int\limits_{S^1{}} dx\, \tr\,[a_k]
\label{4-65b}
\ee
are the spectral invariants of the operator $L$
called global heat kernel coefficients or simply heat invariants.

The first heat kernel coefficient $a_0$ is determined from the initial condition
(\ref{icxx})
and is equal to 
\be
a_0=\II\,.
\ee
The higher-order heat kernel coefficients $a_k, k\ge 1$,
satisfy the following recurrence relations
\cite{avramidi91,avramidi90a,avramidi00b}
\be
\left(1+\frac{1}{k}(x-x')\partial_x\right)a_k=La_{k-1}\,,
\qquad k\ge 1\,.
\ee

A powerful method for calculation of the heat kernel coefficients 
was developed in \cite{avramidi91,avramidi90a,avramidi00b}. 
In the one-dimensional 
case it takes a very simple form
\cite{avramidi00a}.
First of all, we fix the point $x'$.
We introduce the following notation.
For every non-negative integer $n$
we define the functions
\be
|n\rangle=\frac{1}{n!} (x-x')^n;
\ee
we also let, by definition, $|n\rangle=0$ for $n<0$.
Then
\be
D|n\rangle=|n-1\rangle.
\ee
We also define the operator $D^{-1}$ by
\be
D^{-1}f(x)=\int\limits_{x'}^x dy\; f(y)\,;
\ee
then for any non-negative $n$
\be
D^{-1}|n\rangle=|n+1\rangle.
\ee

Next, for every non-negative integer $m$
we define the operators
\be
\langle m| f\rangle=[\partial_x^m f]\,,
\ee
and the matrix elements of a differential operator $L$ by
\be
\left<m\vert L\vert n\right>= \frac{1}{n!}[\partial_x^m L(x-x')^n]\,.
\ee
Then the matrix elements of the 
identity operator are obviously
\be
\langle m|n\rangle=\delta_{mn}\,,
\ee
where $\delta_{mn}$ is the usual Kronecker symbol,
therefore, the matrix elements of the first and the second derivative
have the form
\bea
\left<m|D|n\right> &=&\left<m+1|n\right>=\delta_{n, m+1},
\\
\langle m|D^2|n\rangle &=& \left<m+2|n\right>=\delta_{n, m+2}\,.
\eea
Also, for a function $Q$ for $m\ge n$ we have rather
\be
\left<m|Q|n\right>=\frac{1}{n!}\left[
\partial_x^{m}\left\{Q(x)(x-x')^n\right\}\right]
={m\choose n}Q^{(m-n)}, \qquad m\ge n,
\ee
where
\be
Q^{(n)}=\partial_x^nQ.
\ee
For $m<n$ these matrix elements obviously vanish
\be
\left<m|Q|n\right>= 0, \qquad m\le n-1,
\ee

In general, the matrix elements of a differential operator $L$ of order $p$ vanish
for $m\le n-p-1$,
\be
\left<m|L|n\right>=0 \qquad {\rm for}\ m\le n-p-1.
\ee
For a pseudo-differential (nonlocal) operator it is not so---all matrix
elements
are, in general, non-zero.
For example, 
\be
\langle m|D^{-1}|n\rangle=
\langle m-1|n\rangle= \langle m|n+1\rangle =\delta_{n,{m-1}}.
\ee

The matrix representation of the operators is very convenient
in so far that the products, the powers and the commutators  of the operators 
are given by the product, the powers and the commutators 
of the infinite matrices. For example, two commuting operators must have commuting matrices etc.

By using the above equations we obtain the matrix elements of the Schr\"odinger 
operator (\ref{4-57})
\be
\left<m|L|n\right>
=-\langle m|D^2|n\rangle +\langle m|Q|n\rangle
=-\delta_{n, m+2}+{m\choose n}Q^{(m-n)}\,.
\ee
These matrix elements form an 
infinite matrix
\be
\left(\langle m| L| n\rangle\right)=
\left(
\begin{array}{cccccccc}
Q & 0 & -\II & 0 & 0 & 
\cdots\\
Q^{(1)} & Q & 0 & -\II & 0 & 
\cdots\\
Q^{(2)} & {2\choose 1}Q^{(1)} & Q & 0 & -\II & 
\cdots\\
Q^{(3)} & {3\choose 1}Q^{(2)} & {3\choose 2}Q^{(1)} & Q & 0 & 
\cdots\\
Q^{(4)} & {4\choose 1}Q^{(3)} & {4\choose 2}Q^{(2)} & {4\choose 3}Q^{(1)} & Q &
\cdots\\
\vdots & \vdots & \vdots & \vdots & \vdots & \ddots & 
\end{array}
\right)
\ee

Now, by using the technique developed in \cite{avramidi91} one can
express the coefficients $a_k(x,x')$ in terms of the Taylor series
\be
a_k(x,x')=\sum\limits_{n=0}^\infty \frac{1}{n!}(x-x')^{n}
\left<n|a_k\right>,
\label{4-63}
\ee
where
\bea
\left<n|a_k\right> = [\partial_x^na_k]
&=&\sum_{n_1,\cdots,n_{k-1}\ge 0}\frac{k}{ (k+n)}\cdot
\frac{(k-1)}{ (k-1+n_{k-1})}\cdots\frac{1}{ (1+n_1)}
\qquad\ \,
\nonumber\\[10pt]
& &\times \left<n\vert L\vert n_{k-1}\right>\left<n_{k-1}\vert L\vert n_{k-2}\right>
\cdots
\left<n_1\vert L\vert 0\right>,
\label{4-67x}
\eea
These coefficients are differential polynomials of the 
potential $Q$ evaluated at the point $x'$.

It is important is to note that 
\be
\left<m|L|m+1\right>=0,
\ee
and
\be
\left<m|L|m+2+k\right>=0 \qquad {\rm for}\ k\ge 1.
\ee
and, therefore, 
the summation over $n_i$ in (\ref{4-67x}) is limited from above and 
ranges over
\be
0\le n_1\le n_2+2\le \cdots\le n_{k-1}+2(k-1)\le n+2(k-1).
\ee

By using this technique it is easy to obtain the diagonal values of 
some low-order heat kernel coefficients \cite{avramidi91}
\bea
[a_1]&=&Q,\\
{}[a_2]&=&Q^2-\frac{1}{ 3}Q{''},\\
{}[a_3]&=&Q^3
-\frac{1}{ 2}\left(QQ{''}+Q{''}Q+Q{'}Q{'}\right)
+\frac{1}{ 10}Q^{(4)}.
\eea
The general formula for an arbitrary coefficient $[a_k]$ is presented
in \cite{avramidi00a}.

\section{Leading Derivatives in Heat Kernel Coefficients}
\setcounter{equation}0

The technique described above can be used to analyse the general structure of the
heat kernel coefficients, in particular, to compute
the leading derivative terms in all heat kernel coefficients $[a_k]$.
This has been done in \cite{avramidi90,avramidi91,avramidi00b} for general 
Laplace type operators.
The leading derivatives in the heat kernel coefficients 
for $k\ge 2$ have the following form
\bea
[a_k] &=&
\frac{k!(k-1)!}{(2k-1)!}\left\{\left(-D^2\right)^{k-1}Q
+(2k-1)Q\left(-D^2\right)^{k-2}Q\right\}
\nonumber\\
&&
+O(\partial(QQ))+O(\varepsilon^3).
\eea
Here total derivatives (and commutators) of quadratic terms denoted by $O(\partial(QQ))$ 
and the terms of higher order in $Q$ denoted by
$O(\varepsilon^3)$ are omitted. 

Now, by using the integral
(\ref{456zz})
one can sum up the asymptotic
expansion of the heat kernel diagonal to get
the asymptotic expansion as $\tau, \varepsilon\to 0$
\be
[U(t)] \sim  (4\pi t)^{-1/2}\left\{\II-t\alpha\left(-tD^2\right)Q
+\frac{t^2}{2}Q\alpha\left(-tD^2\right)Q
+O(\partial(QQ))+O(\varepsilon^3)\right\},
\label{52zz}
\ee
where $\alpha(z)$ is exactly the same function defined by
(\ref{458zz}). 
This is an asymptotic expansion as $\tau, \varepsilon\to 0$ but the
parameter $\delta$ does not have to be small,
$\delta\sim 1\,.$

After integrating the heat kernel diagonal all total derivatives
vanish and we obtain then 
the asymptotic expansion as $\tau, \varepsilon\to 0$ of
the heat trace function $\Omega(t)$
\bea
\Omega(t) \sim 2\pi a N 
+\omega(t)
+O(\varepsilon^3)\,,
\label{249}
\eea
where
\bea
\omega(t) &=& -t\int\limits_{S^1} dx\,\tr Q
+\frac{t^2}{2}
\int\limits_{S^1{}} dx\,
\tr Q\alpha\left(-tD^2\right)Q.
\eea
This formula should be compared with the results of Sec. 3.
It can be obtained by taking the limit of large radius $a\to \infty$
in the equation (\ref{317xx}).

Next, by using the equation (\ref{212})
we compute the 
asymptotic expansion of the function $B_q(\lambda)$ as $a\lambda\to -\infty$
and $\varepsilon\to 0$
\be
B_q(\lambda) \sim 2\pi a N (-\lambda)^{q} 
+ b_q(\lambda)
+O(\varepsilon^3),
\ee
where 
\be
b_q(\lambda)=
q(-\lambda)^{q-1}\int\limits_{S^1} dx\,\tr Q
+\frac{{1}}{ {2}}q(q-1)(-\lambda)^{q-2}\int\limits_{S^1{}}dx\,\tr
Qf_{q-2}\left(\frac{D^2}{\lambda}\right)Q,
\ee
and
\be
f_q(z)=\int\limits_0^1
d\xi\,\left(1+\frac{{1-\xi^2}}{ {4}}z\right)^{q}.
\label{fqzz}
\ee

It is easy to see that for positive values of $z$ the function
$f_q(z)$ is an entire function of $q$;
by using eq. (\ref{456zz})
 it can be
represented as a power series
\be
f_q(z)=\sum_{j=0}^\infty 
\frac{\Gamma(q+1)\Gamma(j+1)}{\Gamma(q-j+1)\Gamma(2j+2)}
z^{j}.
\ee
Notice that for non-negative integer values $q=0,1,2,\dots$ this series terminates and is, in fact,
a polynomial of $z$ of order $q$.
One can also compute the asymptotics as $z\to \infty$
\be
f_q(z) = 
\frac{{\Gamma(q+1)\Gamma(q+1)}}{ {\Gamma(2q+2)}}\;z^{q}
+O(z^{q-1}).
\ee

Finally, by using this result the functional determinant,
(\ref{220xx}), takes the form
(within the same accuracy, that is, as an asymptotic series
as $a\lambda\to -\infty$ and $\varepsilon\to 0$)
\bea
\log\Det(L-\lambda)&=&
2\pi a N(-\lambda)^{1/2} + \gamma(\lambda)
+O(\varepsilon^3),
\label{eald}
\eea
where
\bea
\gamma(\lambda) &=&
\frac{1}{2(-\lambda)^{1/2}}\int\limits_{S^1} dx\,\tr Q
-\frac{1}{8(-\lambda)^{3/2}}\int\limits_{S^1{}}dx\,\tr
Qf_{-3/2}\left(\frac{D^2}{\lambda}\right)Q.
\eea
The function $f_{-3/2}$ can be easily computed from (\ref{fqzz});
it has a very simple form
\be
f_{-3/2}(z)=\frac{4}{z+4}.
\ee
Therefore,
\be
\gamma(\lambda) =
\frac{1}{2(-\lambda)^{1/2}}\int\limits_{S^1} dx\,\tr Q
-\frac{1}{4(-\lambda)^{1/2}}\int\limits_{S^1{}}dx\,\tr
Q\left(-D^2-4\lambda\right)^{-1}Q
\ee

The functions $\omega(t)$, $b_q(\lambda)$ and $\gamma(\lambda)$
can be written in the spectral form
as
\bea
\omega(t) &=&
-2\pi a t\, \tr q_0
+2\pi a t^2\sum_{n=1}^\infty |q_{n}|^2 
\alpha\left(\frac{t}{a^2}n^2\right)\,,
\\
b_q(\lambda) &=&
2\pi a q(-\lambda)^{q-1}\tr q_0
+q(q-1)2\pi a (-\lambda)^{q-2}\sum_{n=1}^\infty |q_{n}|^2 
f_{q-2}\left(-\frac{n^2}{\lambda a^2}\right),
\\
\gamma(\lambda)
&=&
\frac{\pi a}{(-\lambda)^{1/2}} \,\tr q_0
-\pi\frac{a^3}{(-\lambda)^{1/2}}\sum_{n=1}^\infty
\frac{|q_n|^2}{n^2-4\lambda a^2}\,,
\eea
where $|q_n|^2=\tr q_n^*q_n$.
We stress here once again that all these results are valid only in the limit
as $a\lambda\to -\infty$ and $\varepsilon\to 0$.
As was noted in Sec. 2 the function $B_q(\lambda)$ and the functional determinant
may be singular at $\lambda\to 0$. Even though the limit $\lambda\to 0$
is not well defined we quote the result
\bea
\gamma(\lambda)&=&
\frac{\pi a}{(-\lambda)^{1/2}}\left\{
\tr q_0
- a^2\sum_{n=1}^\infty|q_n|^2
\frac{1}{n^2}+o(\lambda)\right\}\,,
\eea
which is indeed singular as $\lambda\to 0$.


\section{Equation for Heat Kernel Diagonal}
\setcounter{equation}0

Now,  following \cite{avramidi95}, we derive another recursion 
system for the heat kernel coefficients, 
which gives directly the diagonal heat kernel coefficients $[a_k]$.
It is based on the following purely algebraic lemma.

Let ${\cal L}({\cal H})$ be an algebra of operators 
on some Hilbert space ${\cal H}$. Every operator $Y: {\cal H}\to {\cal H}$ from the algebra
defines the standard action on the algebra itself, $Y: {\cal L}({\cal H})\to {\cal L}({\cal H})$,
by left multiplication $X\to YX$ for any $X\in {\cal L}({\cal H})$;
we will denote this action  by the same symbol $Y$ which should not cause
any confusion. There is also another operator 
 ${\rm Ad}_Y: {\cal L}({\cal H})\to {\cal L}({\cal H})$ defined by the commutator
\be
{\rm Ad}_YX=[Y,X]
\ee
for any $X\in {\cal L}({\cal H})$.

\begin{lemma}
Let $D, Q\in {\cal L}({\cal H})$ be two operators from the algebra ${\cal L}({\cal H})$ and
$L\in {\cal L}({\cal H})$ be an operator defined by
\be
L=-D^2+Q;
\ee
let $U(t)=\exp(-tL)$ be its heat semigroup.

Suppose that the operator ${\rm Ad}_D$ is an injection
and that the image of the operator  ${\rm Ad}_Q$ is a subset of the image
of the operator  ${\rm Ad}_D$, that is, ${\rm Ad}_Q({\cal L}({\cal H}))\subseteq {\rm Ad}_D({\cal L}({\cal H}))$;
then the operator $\hat E: {\cal L}({\cal H})\to {\cal L}({\cal H})$ defined by
\be
{\hat E}=
{\rm Ad}^3_D
-2Q{\rm Ad}_D
-2{\rm Ad}_D Q
+{\rm Ad}_Q{\rm Ad}_D
+{\rm Ad}_D{\rm Ad}_Q
+{\rm Ad}_Q{\rm Ad}^{-1}_{D}{\rm Ad}_Q
\label{4-118}
\ee
is well defined.

Then:
\begin{enumerate}
\item
For any $t\ge 0$,
\be
\left(4\partial_t{\rm Ad}_D
-{\hat E}\right)U{}(t)=0,
\label{4-117}
\ee

\item
and for any non-negative integer $k\ge 0$,
\be
-4{\rm Ad}_DL^{k+1}={\hat E}L^k.
\ee

\end{enumerate}

\end{lemma}

\noindent
{\bf Proof.}
The semigroup satisfies obviously
the equations
\be
\partial_t U{}=-LU{}=-U{}L.
\label{4-111}
\ee
Therefore, we have
\bea
4\partial_t{\rm Ad}_DU{} &=& 
-4DLU+4ULD
\nonumber\\
&=&
-D LU{}
+U{}LD
-3DU{}L
+3LU{}D
\nonumber\\
&=&
D^3U-DQU-UD^3+UQD
\nonumber\\
&&
+3DUD^2-3DUQ-3D^2UD+3QUD
\eea

Now, from the commutativity of the operators $L$ and $U{}(t)$,
\be
[(-D^2+Q),U]=0,
\ee
we have also
\be
[Q,U{}]=[D^2,U{}].
\ee
On the other hand,
\be
[D^2, U{}]=[D,(DU{}+U{}D)],
\ee
and, therefore,
\be
[Q, U{}]=[D,(D U{}+U{}D)].
\ee
This equation can be written as
\be
{\rm Ad}_QU{}={\rm Ad}_D(D U{}+U{}D),
\label{4115xx}
\ee
and, hence,
\be
{\rm Ad}^{-1}_D{\rm Ad}_QU{}=
D U{}+U{}D.
\label{4-115}
\ee
Therefore,
\be
{\rm Ad}_Q{\rm Ad}^{-1}_D{\rm Ad}_QU{}
=QDU+QUD-DUQ-UDQ.
\ee

Next, we compute directly
\bea
{\rm Ad}_D^3U &=& D^3U-3D^2UD+3DUD^2-UD^3,
\\
Q{\rm Ad}_DU &=& QDU-QUD,\\
{\rm Ad}_DQU &=& DQU-QUD,\\
{\rm Ad}_Q{\rm Ad}_D U &=& QDU-QUD-DUQ+UDQ,\\
{\rm Ad}_D{\rm Ad}_QU &=& DQU-DUQ-QUD+UQD.
\eea
By using these results it is easy to show that
\bea
\hat EU
&=&
D^3U-DQU-UD^3+UQD
\nonumber\\
&&
+3DUD^2-3DUQ-3D^2UD+3QUD,
\eea
and, therefore, the heat semigroup satisfies the equation
\be
\left(4\partial_t{\rm Ad}_D
-{\hat E}\right)U{}(t)=0.
\label{4-117xx}
\ee
By expanding this equation in power series in $t$
we also obtain immediately
the commutators 
of the operator $D$ with the powers of the operator $L$
\be
-4{\rm Ad}_DL^{k+1}={\hat E}L^k.
\ee
This is a very important purely algebraical equation that can be proved
also by mathematical induction. 

Now, we apply this lemma to a particular case when
when $D=\partial_x$ is the derivative operator, $Q$
is the operator of multiplication by a matrix-valued function
and
\be
U(t;x,x')=U(t)\delta(x-x')
\ee
is the heat kernel.
Our goal is now to take the equation (\ref{4-117}) in the kernel form, i.e.
to apply it to the delta-function $\delta(x-x')$, and then to compute
its diagonal value $[U(t)]=U(t;x,x)$. 

\begin{corollary}
The heat kernel diagonal $[U(t)]$ of the operator $L=-D^2+Q$
satisfies the equation
\be
\left(4\partial_tD-E\right)[U(t)]=0,
\label{4-107}
\ee
where
\be
E = D^3
-2QD-2D Q
+{\rm Ad}_QD+D{\rm Ad}_Q
+{\rm Ad}_Q{D}^{-1}{\rm Ad}_Q
\label{4-4.46}
\ee
\end{corollary}

\noindent
\paragraph{Remark.}
It should be understood that this operator acts on functions
and not on operators as the operator (\ref{4-118}).
In the scalar case all the commutators vanish and ${E}$ becomes a differential operator
\be
{E} = D^3-2QD-2D Q.
\label{4-4.48}
\ee
The equation (\ref{4-107}) was obtained in \cite{avramidi95} by a completely
different method. Similar equations were obtained in \cite{olmedilla81,bilal94}.

\noindent
\paragraph{Proof.}
First, notice that for every smooth two-point function $f(x,x')$ there holds
\be
(\partial_x+\partial_{x'}) f(x,x')\Big|_{x=x'}=\partial[f],
\label{4-4.39}
\ee
where, as usual $[f]=f(x,x)$ denotes the diagonal value.
Also, for any function $f$ considered as an operator of 
multiplication by this function we have
\be
{\rm Ad}_D f=[D,f]=Df\,,
\ee
and, therefore,
\bea
{\rm Ad}_D U{}(t)\delta(x-x')\Big|_{x=x'}
&=&D[U(t)],
\\
{\rm Ad}^3_D U{}(t)\delta(x-x')\Big|_{x=x'}
&=&D^3 [U(t)],
\\
{\rm Ad}_QU{}(t)\delta(x-x')\Big|_{x=x'}
&=&{\rm Ad}_Q [U(t)],
\\
{\rm Ad}^{-1}_D{\rm Ad}_QU{}(t)\delta(x-x')\Big|_{x=x'}
&=&D^{-1}{\rm Ad}_Q [U(t)]\,.
\eea
By using these equations into eq. (\ref{4-117}) 
we obtain finally the equation (\ref{4-107}).

One should point out that the equation (\ref{4-107}) 
for the heat kernel diagonal is 
a { new nontrivial} equation that expresses deep underlying
symmetry of the one-dimensional spectral problem.
It is this equation that leads to the existence of an
infinite-dimensional completely integrable Hamiltonian system
(Korteweg-De Vries hierarchy).



It is worth noting the following fact. First, we compute
\be
\left(D U{}(t)+U{}(t)D\right)\delta(x-x')\Big|_{x=x'}
=W(t)\,,
\ee
where
\be
W(t)=(\partial_x-\partial_{x'})U(t;x,x')\Big|_{x=x'}.
\label{4-41d}
\ee
Next, we also have
\be
{\rm Ad}_QU{}(t)\delta(x-x')\Big|_{x=x'}
={\rm Ad}_Q [U(t)].
\ee
Therefore, by using eq. (\ref{4115xx}) we obtain
\be
D W(t)={\rm Ad}_Q[U(t)]\,.
\ee

By using the heat kernel expansion (\ref{4-60}) we see that
the function $W$ has the asymptotic expansion
as $t\to 0$
\be
W(t)\sim (4\pi t)^{-1/2}\sum_{k=1}^\infty \frac{(-t)^k}{k!}W_k,
\ee
where
\be
W_k=(\partial_x-\partial_{x'})a_k(x,x')\Big|_{x=x'}\,.
\ee
By comparing this expansion with the eq. (\ref{4-60cx}) we see that
the commutators of the diagonal heat kernel coefficients $[a_k]$ with
the potential $Q$ are also given by the derivative of some
differential polynomials $W_k$, i.e.
\be
D W_k={\rm Ad}_Q[a_k]
\label{4-4.42}
\ee

Note that in the scalar case all commutators with $Q$ vanish
and, therefore, 
\be
W(t)=W_k=0\,.
\ee

\section{Closed Formulas for Spectral Invariants}
\setcounter{equation}0

Substituting the asymptotic expansion of the heat kernel diagonal
(\ref{4-60cx}) into the eq. (\ref{4-107}) we find a direct recursion system
for the diagonal heat kernel coefficients
$[a_k]$
\be
D[a_{k}]=-\frac{k}{ 2(2k-1)}{ E}[a_{k-1}].
\label{4-4.45}
\ee
or, which is equivalent, 
\be
[a_{k}]=\frac{k}{ 2(2k-1)}A[a_{k-1}],
\label{4-4.45a}
\ee
where $A$ is an operator defined by
\be
A=-D^{-1}{E}.
\label{4-4.49a}
\ee
A similar formula  has been found in
\cite{avramidi95,olmedilla81,bilal94}.
Note, that the operator $A$ 
is not a differential operator but a nonlocal pseudo-differential one.
In the scalar case the operator $A$ has a simple form
\be
A=-D^2+2Q+2D^{-1}QD
=-D^2+4Q
-2D^{-1}Q'
\,.
\ee

The recursion system (\ref{4-4.45a}) can be  formally solved:
for $k\ge 1$,
\be
[a_k]=\frac{k!(k-1)!}{ (2k-1)!}A^{k-1}Q.
\label{4-4.49}
\ee
Thus, all diagonal 
heat kernel coefficients $[a_k]$ can be obtained by acting with the 
powers of the operator $A$ on $Q$.

Now, by using this solution for the heat kernel 
coefficients the asymptotic expansion of the heat kernel
diagonal can be summed formally. Indeed, by using eqs.
(\ref{4-60c}) and (\ref{456zz})
we obtain
\bea
[U(t)] & \sim &
(4\pi t)^{-1/2}\left\{\II -t \alpha(tA)Q\right\}\,,
\label{4-126zz}
\eea
where $\alpha(z)$ is the function defined by (\ref{458zz}).
Indeed,  by using the eq. (\ref{dea}) it is easy to see that the heat kernel
diagonal satisfies the eq. (\ref{4-107}).
It is also instructive to compare this result with eq. (\ref{52zz})
obtained by summing the leading derivatives. 
The result (\ref{4-126zz}) goes much further in the sense that
it also sums all powers of the potential $Q$. That is, 
this equation sums all powers of the parameter $\delta$ and $\varepsilon$.
However, it is only valid in the asymptotic limit $\tau\to 0$
(that is, in the limit of the infinite 
radius of the circle, $a\to\infty$). That is why we do not use
the equality sign here. Furthermore, all integrals below over the circle
$S^1$ (of infinite radius) can be
replaced by the integral over the real line $\RR$.
We do not do it since strictly speaking the potential $Q$ is defined
on the circle and we do not assume anything about its behavior
at infinity. 

This gives the spectral function
\bea
\Omega(t) & \sim &
2\pi a N -t \int_{S^1{}}dx\;\tr\alpha(tA)Q,
\label{4-126}
\eea

The closed form (\ref{4-126}) gives then the trace of the 
heat kernel, the zeta-function and
all other spectral functions. 
In particular,
we have for any complex $q$
\be
B_q(\lambda) \sim 2\pi N a(-\lambda)^{q}
+q(-\lambda)^{q-1}\int\limits_{S^1{}} dx\, \tr\,
f_{q-1}\left(-\frac{A}{\lambda}\right)Q,
\ee
where $f_q(z)$ is the function defined by (\ref{fqzz}).
Thus, the functional determinant (\ref{220xx}) takes the form
\bea
\log\Det(L-\lambda) & \sim &
2\pi N a(-\lambda)^{1/2}
+\frac{1}{2(-\lambda)^{1/2}}\int\limits_{S^1{}} dx\, \tr\,
f_{-1/2}\left(-\frac{A}{\lambda}\right)Q.
\eea
The function $f_{-1/2}$ can be easily computed from the definition
(\ref{fqzz}); we get
\be
f_{-1/2}(z)=\frac{2}{\sqrt{z}}\sin^{-1}\left(1+\frac{4}{z}\right)^{-1/2}\,.
\ee



These formal expressions are very useful and provide, for
example, a good framework to obtain the asymptotic expansion of the 
functional determinant as $\lambda\to -\infty$.
Although the limit $\lambda\to 0$ is not well defined, we write
the formal formulas in this case too
\bea
B_q(0) &\sim& \frac{\Gamma(q+1)\Gamma(q)}{\Gamma(2q)}
\int\limits_{S^1{}} dx\, \tr\,
A^{q-1}Q,
\label{4-4.49c}
\\
\log\Det L
&\sim& \frac{\pi}{2}\int\limits_{S^1{}} dx\, \tr\,A^{-1/2}Q.
\eea


\section{Korteweg-de Vries Hierarchy}
\setcounter{equation}0

We describe briefly the formalism of an infinitely-dimensional 
bi-Hamiltonian system
\cite{dickey03}. Let $Q=Q(s)$ be a one-parameter family 
of self-adjoint operators
acting on a Hilbert space ${\cal H}$. Let $H=H(Q)$ be a functional of $Q$. 
Then we define another self-adjoint 
operator $\delta H/\delta Q$ on ${\cal H}$
called the variational derivative of $H$ with respect to $Q$
as follows
\be
\partial_s H
=\Tr\left(\frac{\delta H}{\delta Q}\partial_s Q\right).
\ee

Let $D$ be an anti-self-adjoint operator on the Hilbert space ${\cal H}$.
We define a Poisson bracket on the space of all functionals of $Q$
by
\be
\{F,G\}_{D}=
\Tr\frac{\delta F}{\delta Q}{\rm Ad}_D
\frac{\delta G}{\delta Q}.
\ee
Obviously, it is antisymmetric and satisfies the Jacobi identity, that is,
\bea
&&\{F,G\}_D = -\{G,F\}_{D},
\\
&&\{H,\{F,G\}_D\}_{D}+\{F,\{G,H\}_D\}_{D}+\{G,\{H,F\}_D\}_{D} = 0\,.
\eea

Further, we define a second Poisson bracket
by the operator $\hat E$ (\ref{4-4.46})
\be
\{H,G\}_E=\Tr\frac{\delta H}{\delta Q}\hat E
\frac{\delta G}{\delta Q}.
\label{4-150xx}
\ee
One can show that
this form is indeed antisymmetric and
also satisfies the Jacobi identity.

Now, let $L=L(s)$ be another one-parameter family of
self-adjoint operators on the Hilbert space
defined by $L=-D^2+Q$. Let $U(t)=\exp(-tL)$ be its semigroup
and 
\be
H(t)=-\frac{1}{t}\Tr\exp(-tL)
\ee
be its heat trace; we also define a sequence of functionals
\be
H_k(t)=\partial^k_t H(t).
\label{89xx}
\ee

Then it is easy to show, first, that
\be
\frac{\delta H(t)}{\delta Q}=U(t)\,.
\ee
Next, as we know from (\ref{4-117xx}), 
the heat semigroup satisfies the equation
\be
4{\rm Ad}_D\partial_t U(t)=\hat E U(t),
\ee
where $\hat E$ is the operator defined by (\ref{4-118}).
By multiplying this equation by $U(\tau)$ and taking the trace we obtain
\be
4\Tr U(\tau){\rm Ad}_D\partial_tU(t)=\Tr U(\tau) \hat E U(t),
\ee
which can be written as
\be
\left\{H(\tau), 4\partial_t H(t)\right\}_D
=\left\{H(\tau), H(t)\right\}_{\hat E}.
\ee
The right hand side of this equation is equal to
\bea
\left\{H(\tau), H(t)\right\}_{\hat E}
&=&-\left\{H(t), H(\tau)\right\}_{\hat E}
\nonumber\\
&=&-\left\{H(t), 4\partial_\tau H(\tau)\right\}_D
=\left\{4\partial_\tau H(\tau), H(t)\right\}_D.
\eea
Therefore,
\be
\left\{H(\tau), \partial_t H(t)\right\}_D
=\left\{\partial_\tau H(\tau), H(t)\right\}_D.
\label{814xx}
\ee

Let us define the matrix
\be
M_{kn}=\left\{H_k, H_n\right\}_D\,,
\ee
with $n,k\ge 0$,
where the functionals $H_k$ are defined by (\ref{89xx});
it is obviously, antisymmetric
\be
M_{kn}=-M_{nk}.
\ee
Now, by differentiating the eq. (\ref{814xx}) and setting $\tau=t$ 
we see that this 
matrix satisfies the equation
\be
M_{nk}=M_{n+1,k-1}\,.
\ee
Therefore, the matrix $M$ vanishes on the main diagonal
and on the next to the main diagonal
\be
\qquad
M_{nn}=M_{n,n+1}=0\,.
\ee
Now, we show by induction that it vanishes on all diagonals;
we have for any $n,k\ge 0$
\be
M_{n, n+2k}=M_{n+1,n+2k-1}=\cdots
=M_{n+k,n+k}=0,
\ee
and
\be
M_{n,n+2k+1}=M_{n+1,n+2k}=\cdots
=M_{n+k, n+k+1}=0.
\ee
This proves that this matrix is equal to zero, $M_{kn}=0$. That is,
the derivatives of the function $H(t)$ are all in involution
\be
\left\{H_k,H_n\right\}_D=0
\ee
for any $k,n\ge 0$.

Next, we define an hierarchy of Hamiltonian systems
(that we call a {\it generalized KdV hierarchy})
\be
\partial_s Q={\rm Ad}_D H_k,
\ee
with the parameter $t$ in $H_k(t)$ being fixed here.
Then for any functional $\Phi$ of the operator $Q$ we have
\be
\partial_s\Phi 
=\Tr\frac{\delta \Phi}{\delta Q}\partial_s Q
=\Tr\frac{\delta \Phi}{\delta Q}{\rm Ad}_D
\frac{\delta H_k}{\delta Q}
=\{\Phi, H_k\}_D\,.
\ee
Therefore, a functional $\Phi$ is an integral of motion 
of the Hamiltonian system if and only if
its Possion bracket with the Hamiltonian $H_k$ vanishes
(that is, it is in involution with the Hamiltonian).
Thus, all functionals $H_k$ are integrals of motion of the
whole hierarchy of Hamiltonian systems, that is, for any $n,k$,
\be
\partial_s H_n=0.
\ee


A special motivation for the study of the one-dimensional heat kernel
is its relation to the Korteweg-de Vries (KdV) hierarchy.
We consider a second-order differential operator of the form
$L=-D^2+Q$;
to be specific, we assume that the potential $Q$ is a real symmetric
matrix.

We will need to study the deformation of spectral invariants under the variation
of the potential $Q$.
More specifically we consider a one parameter family of operators
$L(s)=-D^2+Q(s)$, where $s$ is a real parameter.
Then we have
\be
\partial_{s}\Theta(t)=
\partial_{s}\Tr \exp(-tL)=-t\Tr\left[\partial_s Q\exp(-tL)\right]\,.
\ee
This means that
\be
\frac{\delta \Theta(t)}{\delta Q}=-t[U(t)].
\ee
Expanding both sides of this equation in the asymptotic series as $t\to 0$, we
see that
\be
\partial_s A_k=k\int\limits_{S^1{}} dx\;\tr \partial_s Q[a_{k-1}],
\ee
where $A_k$ are the global heat kernel coefficients (\ref{4-65b})
of the operator $L$
and $[a_k]$ are the diagonal local heat kernel coefficients introduced 
in the Sec. 4.
Therefore,
\be
\frac{\delta A_k}{\delta Q}=k[a_{k-1}].
\label{440xx}
\ee


Now, we rescale the sequence of global heat invariants $A_{k}$
and define a new sequence $I_k$ by
\be
I_{k} = (-1)^{k}\frac{(2k)!}{ k!(k+1)!}A_{k+1}.
\ee
Then by using eqs. (\ref{4-4.49}) and (\ref{4-4.49a})
\be
\frac{\delta I_k}{\delta Q}
=-2(D^{-1}E)^{k-1}Q,
\label{4-126a}
\ee
and, therefore,
\be
D\frac{\delta I_k}{ \delta Q}
=E\frac{\delta I_{k-1}}{\delta Q},
\ee

These functionals define the KdV hierarchy
\be
\frac{\partial Q}{\partial s}
=D\frac{\delta I_k}{\delta Q}, \qquad k=1,2,\dots .
\label{4-4.50}
\ee
This system is an infinitely-dimensional 
bi-Hamiltonian system. 
We define two Poisson brackets
\bea
\{H,G\}_{D} &=&
\int\limits_{S^1{}} dx\, \tr\,
\frac{\delta H}{\delta Q}D
\frac{\delta G}{\delta Q},
\\
\{H,G\}_E &=& \int\limits_{S^1{}} dx\, \tr\,
\frac{\delta H}{\delta Q}E
\frac{\delta G}{\delta Q},
\label{4-150}
\eea
where $E$ is the operator defined by (4-4.48).
Now by using (\ref{4-4.46}) one can show that
this form is indeed
antisymmetric
in spite of the fact, that the operator $E$ is not 
anti-self-adjoint, in general.
In the scalar case the operator $E$ given by 
(\ref{4-4.48}) is anti-self-adjoint and
the corresponding form $\{F,G\}_E$ (\ref{4-150}) 
is antisymmetric automatically.
This means that the Poisson brackets are related by
\be
\{I_{n},I_{k}\}_D=\{I_{n},I_{k-1}\}_E.
\ee
Now, exactly as above, 
this enables one to show that
\begin{itemize}
\item[i)]
all functionals $I_k$ are in involution, that is, for any $n,k$,
\be
\{I_{n},I_{k}\}_D=\{I_{n},I_{k}\}_E=0,
\ee
\item[ii)]
and, therefore, are integrals of motion, that is, for any $n$,
\be
\partial_s I_n=0.
\ee
\end{itemize}


The generalization of this scheme further (to partial differential operators
on manifolds, pseudo-differential operators, discrete operators, etc) is an
interesting and intriguing problem related to the whole area
of spectral geometry and isospectral deformations.
What one has to do is to find two anti-self-adjoint 
operators ${\cal D}$ and ${\cal E}$, such that the heat kernel diagonal
satisfies the equation
\be
(4{\cal D}\partial_t-{\cal E})[U(t)]=0.
\ee
If such operators are found and ${\cal D}$ satisfies additionally the
Jacobi identity, then the whole construction can be carried out to
obtain a completely integrable infinitely dimensional Hamiltonian
system.



\end{document}